\documentclass{article}
\usepackage{spconf,amsmath,graphicx}
\usepackage[moderate]{savetrees}
\usepackage{epigraph}
\usepackage{amssymb}
\usepackage{pifont}
\usepackage{multirow}
\usepackage{footmisc}
\usepackage{subcaption}
\usepackage{makecell}
\usepackage{listings}
\usepackage{xcolor}
\usepackage{xspace}
\usepackage{stix}
\usepackage{dirtytalk}
\usepackage{amssymb}
\usepackage{amsmath}
\usepackage{amsfonts}
\usepackage{booktabs}
\usepackage{algorithm}
\usepackage{algpseudocode}

\usepackage{hyperref}
\hypersetup{
    colorlinks=true,
    linkcolor=purple,
    filecolor=magenta,      
    urlcolor=purple,
}

\usepackage[numbers,sort]{natbib}
\newcommand{\datasetname}{\texttt{VoxMovies}}
\newcommand{\VoxCelebDomain}{{\it D-I}}
\newcommand{\VoxMovieDomain}{{\it D-M}}
\newcommand{\VoxMovieDomainsamemovie}{{\it D-M (same)}}
\newcommand{\VoxMovieDomaindifferentmovie}{{ \it D-M (diff)}}
\newcommand{\CrossDomain}{\VoxCelebDomain, \VoxMovieDomain}
\newcommand{\VoxMovieTrain}{\datasetname-(Train)}
\newcommand{\VoxMovieTest}{\datasetname-(Test)}

\usepackage[utf8]{inputenc}
\usepackage[title]{appendix}

\definecolor{codegreen}{rgb}{0,0.6,0}
\definecolor{codegray}{rgb}{0.5,0.5,0.5}
\definecolor{codepurple}{rgb}{0.58,0,0.82}
\definecolor{backcolour}{rgb}{0.95,0.95,0.92}
 
\lstdefinestyle{mystyle}{
    backgroundcolor=\color{backcolour},   
    commentstyle=\color{codegreen},
    keywordstyle=\color{magenta},
    numberstyle=\tiny\color{codegray},
    stringstyle=\color{codepurple},
    basicstyle=\ttfamily\footnotesize,
    breakatwhitespace=false,         
    breaklines=true,                 
    captionpos=b,                    
    keepspaces=true,                 
    numbers=left,                    
    numbersep=5pt,                  
    showspaces=false,                
    showstringspaces=false,
    showtabs=false,                  
    tabsize=2,
    basicstyle=\fontsize{7}{8}\ttfamily
}
 
\newcommand{\newpara}[1]{\vspace{8pt}\noindent\textbf{#1}}
\usepackage{array}
\newcommand{\PreserveBackslash}[1]{\let\temp=\\#1\let\\=\temp}
\newcolumntype{C}[1]{>{\PreserveBackslash\centering}p{#1}}
\newcolumntype{R}[1]{>{\PreserveBackslash\raggedleft}p{#1}}
\newcolumntype{L}[1]{>{\PreserveBackslash\raggedright}p{#1}}
\def\pretabcapt{\vspace{-7pt}}
\title{Playing a Part: Speaker Verification at the Movies}
\name{Andrew Brown$^{1}$*, Jaesung Huh$^{1}$*, Arsha Nagrani$^{1,\dagger}$*, Joon Son Chung$^{2}$, Andrew Zisserman$^{1}$\thanks{\hspace{-10pt}* These authors contributed equally to this work.}\thanks{\hspace{-10pt}$\dagger$ Now at Google Research.}}
\address{$^{1}$Visual Geometry Group, University of Oxford, UK \\
$^{2}$Naver Corporation, South Korea \\
\small{\texttt{\{abrown,jaesung,arsha,joon,az\}@robots.ox.ac.uk}} \\
}
\begin{document}
\ninept
\maketitle
\begin{abstract}

\end{abstract}
The goal of this work is to investigate the performance of popular speaker recognition models on speech segments from movies, where often actors intentionally disguise their voice to play a character. We make the following three contributions: (i) We collect a novel, challenging speaker recognition dataset called \datasetname, with speech for 856 identities from almost 4000 movie clips.
\datasetname\ contains utterances with varying emotion, accents and background noise, and therefore comprises an entirely different domain to the interview-style, emotionally calm utterances in current speaker recognition datasets such as VoxCeleb; (ii) We provide a number of domain adaptation evaluation sets, and  benchmark the performance of state-of-the-art speaker recognition models on these evaluation pairs. We demonstrate that both speaker verification and identification performance drops steeply on this new data, showing the challenge in transferring models across domains; and finally (iii) We show that simple domain adaptation paradigms improve performance, but there is still large room for improvement. 


\section{Introduction}
\label{sec:intro}
\begin{flushright}
\textit{All the world's a stage, and all the men and women merely players; They have their exits and their entrances, and one man in his time plays many parts.} \\
- W. Shakespeare, As You Like It
\end{flushright}

After hearing the actor Steve Martin's smooth American accent on the Ellen show, is it possible to recognise that he is the voice behind `Inspector Jacques Clouseau', the comical French character in the movie `Pink Panther', without access to his visual appearance? Or recognise Anne Hathaway's voice in the movie `Les Misérables', where she plays `Fantine', singing through tears about her sadness and desperation? In this paper we investigate the challenging task of speaker recognition for actors from two different domains -- the first being when they are speaking {\em `naturally'} in interviews, and second while {\em playing a part} in a movie, where they may be intentionally modifying their voice in order to play the role of a character or show emotion.  

While recent years have shown great successes in speaker recognition~\cite{snyder2018x, garcia2020jhu, wan2018generalized, chung2020defence}, these successes have been reliant on the collection of large, labelled datasets such as VoxCeleb~\cite{Nagrani17,Chung18a,nagrani2020voxceleb} and others~\cite{mclaren2016speakers,fan2020cn}. The VoxCeleb datasets, while valuable, have been collected \textit{entirely from interviews} of celebrities in YouTube videos and are limited in terms of linguistic content (celebrities mostly speak about their professions~\cite{Nagrani20d}), emotion, and background noise. In contrast, movies contain speech covering emotions such as anger, sadness, assertiveness, and fright, and varied background conditions -- imagine the shouting in a violent scene from an action movie, or a romantic scene of reconciliation in a romcom. As we show in this paper, models trained on VoxCeleb, when applied to a novel domain such as speech in movies, suffer from significant degradation in performance. 
In order to accurately measure this, there is a compelling need for real-world datasets and evaluation sets across these domains. Collecting and annotating datasets for every new domain encountered in the real-world, however can be an extremely expensive and time-consuming process. We introduce a scalable method to automatically generate data in a new domain (movies), and investigate the performance of state-of-the-art speaker recognition models on this data, where actors are intentionally disguising their voice. Being able to detect human identity under such conditions of spoofing is valuable for security and authentication~\cite{cai2017countermeasures, chen2017resnet, das2019long}, and as shown by psychology studies~\cite{reich1979effects, hirson1993glottal}, is a challenging task even for humans.

\begin{figure*}[h]
\begin{center}
\includegraphics[width=\linewidth]{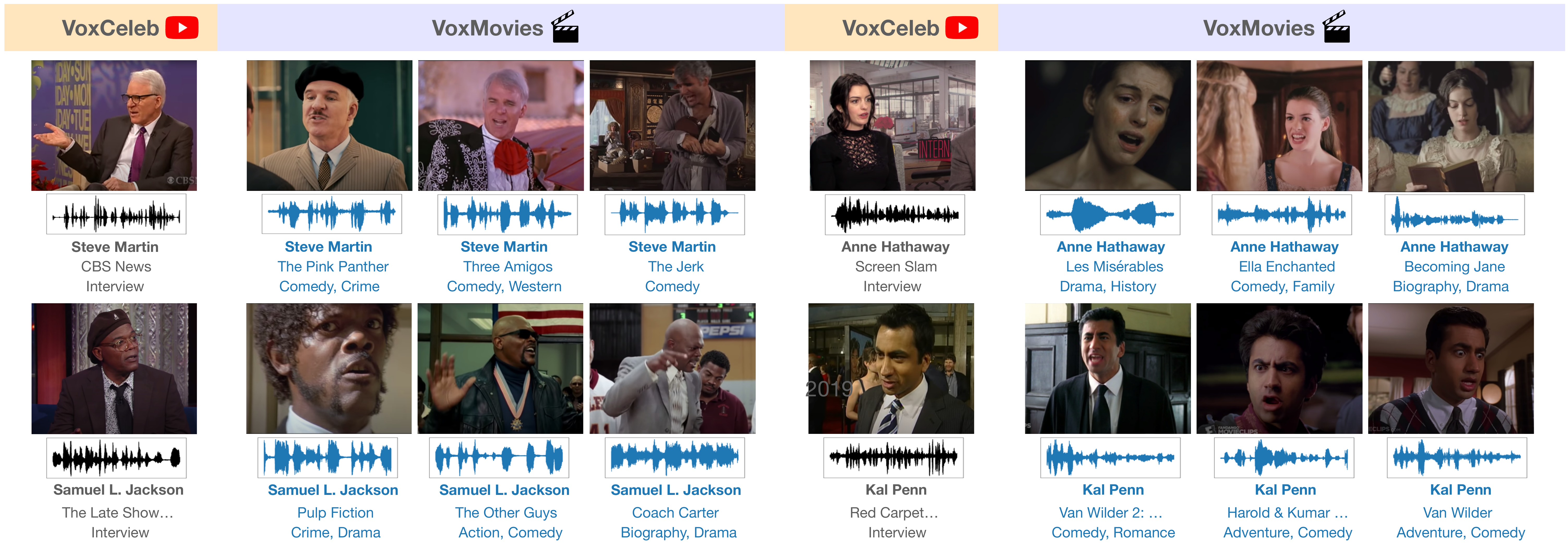}
\end{center}
\vspace{-15pt}
  \caption{\small\textbf{Domain Gap between VoxCeleb and \datasetname:} Unlike VoxCeleb (left in each panel), which consists of utterances from interviews, \datasetname\ is sourced from movies of different genres (right in each panel). While VoxCeleb features utterances largely with calm, unvaried emotions, \datasetname\ has challenging emotion and background noise. We show only a single frame for each utterance, below which is the name and genre of the film/video. \datasetname\ contains 24 utterances from 5 movies for Anne Hathaway, including singing in the musical, `Les Misérables', reading in an English accent in `Becoming Jane', and arguing heatedly in an American accent in `Ella Enchanted'.  Samuel L. Jackson has 59 utterances from 14 movies, including his famous, assertive speech in `Pulp Fiction' and bragging as an arrogant policeman in `The Other Guys'. }
\label{fig:teaser}
\vspace{-12pt}
\end{figure*}

In order to encourage research in domain adaptation for speaker recognition, we make the following three contributions: (i) We collect a novel speaker recognition dataset called \datasetname, from 3,792 popular movie clips uploaded to YouTube. Our dataset consists of almost 9,000 utterances from 856 identities that appear in the VoxCeleb datasets, and contains challenging emotional, linguistic and channel variation (Fig.\ \ref{fig:teaser}); 
(ii) We provide a number of domain adaptation evaluation sets, and  benchmark the performance of state-of-the-art speaker recognition models on these evaluation pairs. We demonstrate that performance drops steeply on this new data for both speaker verification \textit{and} identification, showing the challenge in transferring models across domains (from interviews to movies). We also investigate performance on positive pairs sampled across different movies, and reveal further performance drops; and finally (iii) We demonstrate that domain adaptation approaches added on top of already trained models improve performance, but there is still a severe degradation.

\datasetname\ has  been used to create a challenging test set for the VoxCeleb Speaker Recognition Challenge~\cite{nagrani2020voxsrc} (VoxSRC2020)\footnote{http://www.robots.ox.ac.uk/~vgg/data/voxceleb/competition2020.html\label{refnote}}. Data: \url{https://www.robots.ox.ac.uk/~vgg/data/voxmovies/}.

\vspace{-2mm}
\section{Related Work}
\label{sec:related_work}
\vspace{-1mm}

Due to the reliance of machine learning on large, labelled datasets, domain adaptation has become popular in fields such as computer vision~\cite{tzeng2017adversarial, hoffman2018cycada}, text classification~\cite{liu2017adversarial, zhang2017aspect}, speech enhancement~\cite{liao2018noise, meng2018adversarial} and speaker verification~\cite{garcia2014supervised,garcia2014unsupervised,garcia2014improving}. Recent interest in domain adaptation for speaker recognition has largely focused on boosting the performance on datasets such as NIST-SRE16 or Speakers in the Wild (SITW)~\cite{mclaren2015speakers}, which does not have a large training set. In this case, most methods train on VoxCeleb~\cite{Nagrani17, Chung18a} and evaluate on these evaluation sets. 

Recent methods focus on adversarial training techniques, with \cite{wang2018unsupervised} introducing domain adversarial training that exploits domain prediction, and~\cite{luu2020channel} proposing a channel adversarial training method by adding a channel discriminator to mitigate the domain mismatch problem by using the video labels in VoxCeleb, with both works~\cite{luu2020channel,wang2018unsupervised} using gradient reversal layers. 
~\cite{rohdin2019speaker} proposes an end-to-end domain adversarial training method adopting the architecture of Wasserstein Generative Adversarial Network (GAN)~\cite{arjovsky2017wasserstein} with adversarial domain loss and cross-entropy loss. It significantly improves the performance on language adaptation. \cite{bhattacharya2019adapting} also introduces a domain adaptation technique to new language or recording conditions by using a Domain Adversarial Neural Speaker
Embeddings (DANSE) model which contains a 1-dimensional self-attentive residual block. ~\cite{bhattacharya2019generative} explores various GANs to make the domain discriminator unable to distinguish whether embeddings are from the source or target domain.   \cite{kang2020domain} exploits Model-Agnostic Meta-Learning (MAML), projecting speaker representations to a generalized embedding space and achieves better results on CNCeleb dataset~\cite{fan2020cn}. Other research methods train channel-invariant, noise-robust speaker recognition models which can help improve the overall performance of the model in diverse domains. \cite{zhou2019training} shows a multi-task learning method that trains both a speaker recognition network and a discriminator that distinguishes types of noise in speaker representations. \cite{chung2019delving} also tackles a similar problem by using an environment network and a KL-divergence based confusion loss to learn speaker-discriminative, environment-invariant representations.

Unlike existing works, we provide novel domain data for the \textit{same identities as in VoxCeleb}, allowing us to investigate both cross-domain speaker verification, where pairs have one segment from either domain, and cross-domain identification, where speaker identification models are tested on a new, previously unseen domain. Both have not been previously possible. We focus on the domain adaptation from interviews to movies for male and female actors (explored for faces in~\cite{nagrani2018benedict}). 
We propose and benchmark on such evaluations conditions, and additionally on several within-domain verification tasks. 

\vspace{-2.5mm}
\section{Cross-Domain Data}
\vspace{-1mm}
\label{sec:dataset}

Our goal in this work is to investigate the effects of cross-domain speaker verification in movies. The domain change we focus on here is from YouTube interviews (domain \VoxCelebDomain), to speech in different genres of movies (domain \VoxMovieDomain). The data for the two different domains are sourced as follows: \\
\noindent\textbf{\VoxCelebDomain: Interviews from VoxCeleb~\cite{Nagrani17, Chung18a}} These datasets are sourced solely from interviews uploaded to YouTube. These are mainly in studio, outdoor or red-carpet locations. In turn, and due to the often \textit{professional} context, voices are mainly calm and rarely show any strong emotion. These utterances are degraded with real world noise that would be expected from these environments, such as background chatter or laughter. \\
\noindent\textbf{\VoxMovieDomain: Movies from CMD~\cite{Bain20} (VoxMovies)} For the second domain, we curate a dataset of speech from movies, called \datasetname, which consists of 8,905 utterances for 856 different identities, sourced from 3,792 video clips from 1,452 movies. These movies cover a range of genres (see Figure~\ref{fig:histograms}). The utterances in \datasetname\ are sourced from the Condensed Movies dataset (CMD)~\cite{Bain20}, which covers the \textit{key scenes} from movies. The distinctive change of domain can be seen in the following characteristics of  \datasetname: \\
\noindent\textbf{(1) Emotion:} In line with different movie genres, the utterances cover emotions such as anger, sadness, assertiveness, and fright. Furthermore, the videos in CMD represent scenes that are integral to the story-line and the different character developments, such as a fight between two main characters, or when they make up later in the film. Hence the utterances in the dataset often capture the most emotional parts of each movie. \\
\noindent\textbf{(2) Background noise:} Each key scene in the CMD dataset covers many different settings, from a loud basketball stadium in \textit{Coach Carter}, to an 18th century gathering in \textit{Becoming Jane} (see Figure~\ref{fig:teaser}). This represents a far more varied set of degradation for each of the utterances. Also, there is often background music.

Importantly, in \datasetname\ this variety of emotion and background noise is seen both within and across different identities. Firstly this is because on average each identity has utterances from 2.7 different movies (see Table~\ref{dataset_stats}), and these movies are likely to be of different genres. Secondly, the videos in this dataset show the important character arcs of these identities within each movie, where they show different emotions at different points in the story-line. Examples and further details can be found in Figure~\ref{fig:teaser}.

Note that the utterances in \datasetname\ are all from identities that are represented in VoxCeleb1 and VoxCeleb2.  We create an evaluation set, \VoxMovieTest, featuring identities from VoxCeleb1. We also provide a small amount of \datasetname\ data for training domain adaptation methods, \VoxMovieTrain, using a subset of the identities in the VoxCeleb2 dev set.  There is no identity overlap between these partitions. Statistics are shown in Table~\ref{dataset_stats}. 
\vspace{-4mm}
\section{Dataset Collection Pipeline}
\vspace{-1mm}
\label{sec:dataset_colletion}
Our dataset collection pipeline is similar to the one used to collect the VoxCeleb datasets, albeit applied to YouTube clips of movie scenes from the Condensed Movies Dataset (CMD)~\cite{Bain20}, described below.  

\begin{figure}[t!]

\centering
\includegraphics[width=.23\textwidth]{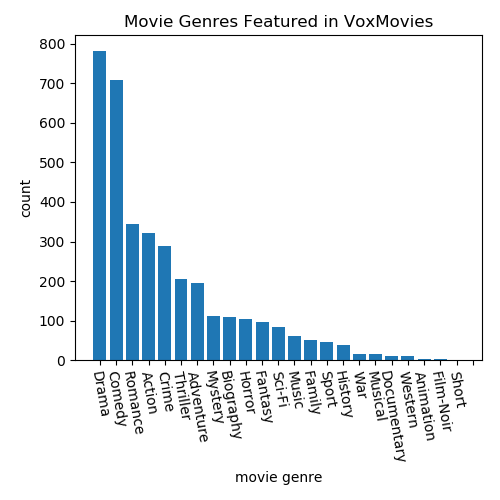}\hfill
\includegraphics[width=.232\textwidth]{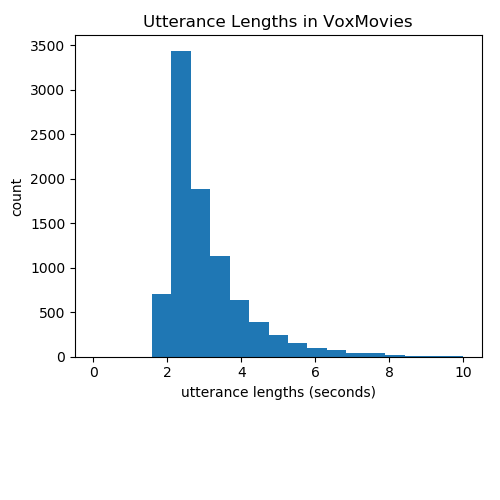}\hfill
\vspace{-8pt}
\caption{\small\textbf{Statistics of the \datasetname\ dataset.} (a) The different movie genres that the utterances in \datasetname\ are sourced from, and (b) the distribution of utterance lengths. The minimum length is 2 seconds by design choice. 
}
\label{fig:histograms}
\vspace{-2mm}
\end{figure}

\begin{table}[]
\footnotesize
\centering
\begin{tabular}{cccccc}
\toprule
\textbf{Partition} & \textbf{\begin{tabular}[c]{@{}c@{}} ID Source\end{tabular}} & \textbf{IDs} & \textbf{\# Utter.} & \textbf{\begin{tabular}[c]{@{}c@{}}\ Clips \\ / ID\end{tabular}} & \textbf{\begin{tabular}[c]{@{}c@{}} Movies  \\ / ID\end{tabular}} \\ 
\midrule 

\VoxMovieTest              & VoxCeleb1                                                           & 485                 & 4,943               & 5.1                                                                    & 2.7                                                                     \\ 
\VoxMovieTrain               & VoxCeleb2                                                           & 371                 & 3,962               & 5.0                                                                    &          2.5   
\\ \bottomrule
\end{tabular}
\pretabcapt
\caption{\small{\textbf{Dataset Statistics for \datasetname.} The utterances are sourced from the Condensed Movies Dataset~\cite{Bain20}, which contains clips of key scenes from movies. Identities (IDs) overlap with the VoxCeleb1 (test) and VoxCeleb2 (dev) datasets. 
}}
\label{dataset_stats}
\vspace{-2mm}
\end{table}

\noindent
\textbf{Condensed Movies Dataset.}
CMD~\cite{Bain20} consists of key scenes from over 3K movies, totalling 1,270 hours. Provided alongside the dataset are cast lists for each of the featured movies, face-tracks for each of the clips, and face embeddings for discriminating identity for each of these face-tracks. The cast lists give the names of people who are likely to appear in the clip. 
Our three stage method for collecting \datasetname\ is as follows: 

\noindent 
\textbf{Stage 1. Sourcing candidate names.} We compute the intersection between the VoxCeleb1 and VoxCeleb2 identities and the CMD cast lists.  

\noindent 
\textbf{Stage 2. Face verification.} In this stage, face-tracks from the CMD dataset are classified as to whether they depict any of the candidate names from the previous stage using a three step process. (1) Example face images for each of the candidate names for the train and test set are sourced from the VGGFace2~\cite{Cao18} and VGGFace1~\cite{Parkhi15} datasets (these face datasets contain the same identities as the VoxCeleb2 and VoxCeleb1 datasets respectively). (2) 256D Embeddings are taken from the final hidden layer of a SE-Net50 architecture CNN for each of the face images, trained for discriminating face-identity using the VGGFace2 dataset~\cite{Cao18}. For each identity, these embeddings are average pooled across all example face-images and L2 normalised, leaving one embedding per identity. (3) Verification is performed by computing the cosine similarity between the identity embeddings and the face-track embeddings from CMD. Any face-track with a similarity to an identity embedding above a high threshold is assigned that name.

\noindent 
\textbf{Stage 3. Active speaker verification.} Here we identify which face-tracks from the previous stage are speaking. This is performed using an audio-visual synchronisation network~\cite{chung2019perfect}, which predicts the correlation between the audio track and the motion of the mouth, and outputs a synchronisation confidence score for each 5 frame window. A window with confidence above a high threshold is classified as speaking. Face-tracks with a minimum of 2 seconds of consecutive speech are kept, and the audio track is clipped to the speaking period. See Figure~\ref{fig:histograms} for a histogram of segment lengths.

\noindent
\textbf{Discussion.} A manual check of \VoxMovieTest\ reveals that the automated process has a precision of $>99\%$. The $<1\%$ false positive face verifications, or false positive active speaker verifications are manually removed. The high thresholds for stages 2 and 3 are chosen as 0.55 and 0.13, respectively. These values (both cosine similarity) achieved high precision on a manually labelled validation set. As shown in Table~\ref{dataset_stats}, \VoxMovieTest\ has slightly more utterances than \VoxMovieTrain. This is due to the fact that a larger proportion of the identities in VoxCeleb1 are well-known actors than in VoxCeleb2, and hence they appeared in more of the movies in CMD. Impressively, on average, each identity has utterances from 2.7 different movies, with Robert DeNiro appearing in 25 movies.

\vspace{-2mm}
\section{Experiments}
\vspace{-1mm}
\label{sec:experiments}
\begin{table}[]
\footnotesize
\centering
\begin{tabular}{cccccc}
\toprule
\begin{tabular}[c]{@{}c@{}}\textbf{Eval.} \\ \textbf{set}\end{tabular} & \begin{tabular}[c]{@{}c@{}}\textbf{Positives} \\ \textbf{Source}\end{tabular} & \begin{tabular}[c]{@{}c@{}}\textbf{Negatives} \\ \textbf{Source}\end{tabular} &\begin{tabular}[c]{@{}c@{}}\textbf{\# Utter.}\end{tabular}  & \begin{tabular}[c]{@{}c@{}}\textbf{\# Positive} \\ \textbf{Pairs}\end{tabular} & \begin{tabular}[c]{@{}c@{}}\textbf{\# Negative} \\ \textbf{Pairs}\end{tabular} \\ 
\midrule 


  E-1    &   \VoxMovieDomainsamemovie\  & \VoxMovieDomain\  &     20,572                                                      &    10,286            &   10,286  \\
 E-2      &  \CrossDomain\ &  \CrossDomain\  &    46,578                                                        &      23,289            &   23,289  \\
  E-3       &  \CrossDomain\ &   \VoxCelebDomain\    &    46,804                                                        &      23,402            &   23,402   \\
 E-4     & \CrossDomain\  &  \VoxMovieDomain\  &    46,866                                                        &      23,433            &   23,433  \\
 E-5    & \VoxMovieDomaindifferentmovie\  & \VoxMovieDomain\  &      41,090                                                      &       20,545           &   20,545  \\

\bottomrule
\end{tabular}
\pretabcapt
\caption{\small{\textbf{Statistics for the different evaluation sets:} Our evaluation sets are sourced from two different domains, interview material (\VoxCelebDomain) and movie material (\VoxMovieDomain). Key: \textbf{Utter.}:  Utterances., { \textbf \VoxMovieDomainsamemovie:} Segments in a pair are sourced from the same movie, { \textbf \VoxMovieDomaindifferentmovie:} Segments sourced from different movies. }}
\label{eval_stats}
\vspace{-3mm}
\end{table}

\subsection{Evaluation Tasks}
Our goal is to determine the performance of state-of-the-art models trained on the VoxCeleb2 dev set for speaker recognition performance on data from the movie domain (\VoxMovieDomain). 

\noindent\textbf{Verification:}
We use the VoxCeleb1 and \VoxMovieTest\ datasets for evaluation, as these sets have no overlap with the identities in the VoxCeleb2 dev set (which is used to train all baselines). Given the two domains (\VoxCelebDomain\ and \VoxMovieDomain), we note that there are three different ways that pairs can be sourced for evaluation: pairs can be sourced entirely from \VoxCelebDomain, entirely from \VoxMovieDomain, or from both, where one utterance is from \VoxCelebDomain\ and one is from \VoxMovieDomain. For positive pairs sourced from \VoxMovieDomain, we can add the further constraint that both utterances must come from the same movie (\VoxMovieDomainsamemovie) or from different movies (\VoxMovieDomaindifferentmovie). We use these options to create 5 challenging evaluation sets E1-E5 (Table~\ref{eval_stats}), of increasing difficulty. More details are provided in Section~\ref{sec:Results}.

\noindent\textbf{Identification:}
We also demonstrate domain mismatch from \datasetname\ and VoxCeleb, with speaker identification. Here we add a single linear layer on baseline models trained on VoxCeleb2 dev set for verification. The task is to fine-tune this layer with the cross-entropy loss using utterances from 485 speakers in the VoxCeleb1 dataset (using only the training data from the identification split, see Table 5 in~\cite{Nagrani17}). We then test performance using the VoxCeleb1 identification test data (in-domain test set) and the utterances from \datasetname\ (out of domain test set). 

\noindent\textbf{Evaluation Metrics:}
For verification, we report equal error rate (\textbf{EER- 
$\%$})\  and minimum detection cost \textbf{(minDCF)} with $C_{miss}=1$, $C_{fa}=1$ and $P_{target}=0.05$. The formula of minimum detection cost function is the same as the one used in NIST SRE~\cite{nist2018} and the VoxSRC2020 evaluations. For identification, we report top 1 and top 5 \% accuracy. 
\begin{table}[!tb]

\scriptsize
\setlength{\tabcolsep}{4pt}
\renewcommand\arraystretch{1.2}
\centering
\begin{tabular}{ l  | cc | cc  | cc  | cc }
\toprule
  \textbf{Eval set} &\multicolumn{2} {c|} {\begin{tabular}[c]{@{}c@{}}\textbf{I-vec.} \textbf{\cite{dehak2010front}}\end{tabular}} & \multicolumn{2} {c|}{\begin{tabular}[c]{@{}c@{}}\textbf{X-vec.} \textbf{\cite{snyder2018x}}\end{tabular}} & \multicolumn{2} {c|} {\textbf{\begin{tabular}[c]{@{}c@{}}\textbf{Thin-R34~\cite{cai2018exploring}} \end{tabular}}} & \multicolumn{2} {c} {\begin{tabular}[c]{@{}c@{}}\textbf{Thick-R34~\cite{heo2020clova}}\end{tabular}}  \\
  & \textbf{EER} & \textbf{DCF} & \textbf{EER} & \textbf{DCF} & \textbf{EER} & \textbf{DCF} & \textbf{EER} & \textbf{DCF} \\
\midrule
\multicolumn{1}{l|}{\begin{tabular}[c]{@{}l@{}}VoxCb1$\dagger$\end{tabular}} & 5.53 & 0.336 & 3.30 & 0.220& 2.05& 0.166 & 1.05&0.084\\
\multicolumn{1}{l|}{\begin{tabular}[c]{@{}l@{}}VoxCb1-H$\dagger$\end{tabular}} & 9.13 & 0.467 & 5.82 & 0.352& 4.37&0.283& 2.39&0.154\\
\multicolumn{1}{l|}{\begin{tabular}[c]{@{}l@{}}VoxCb1-E$\dagger$\end{tabular}} & 5.55&0.338& 3.35 & 0.221 & 2.27&0.164& 1.22&0.086\\

\midrule
\multicolumn{1}{l|}{\begin{tabular}[c]{@{}l@{}}VoxSRC20*\end{tabular}} & 11.07&0.566& 8.22& 0.450 & 6.35& 0.374 & 3.79 & 0.213\\
\midrule
E-1          & 16.6&0.727& 12.92 & 0.665& 9.72&0.562& 6.09&0.365\\
E-2          & 17.91& 0.822& 14.75 & 0.712 & 10.58& 0.610 & 7.40&0.423\\
E-3         & 18.91 & 0.917 & 13.58 & 0.806 & 10.58 &0.666& 7.50& 0.484\\
E-4           & 19.83& 0.872& 21.56 & 0.845& 12.52 & 0.737 & 9.23&0.579\\
E-5          & 21.5&0.913& 17.97 & 0.861& 14.11&0.760& 10.47&0.574\\
 \bottomrule
\end{tabular} 
\pretabcapt
\caption{\small{\textbf{Baseline results on various \datasetname\ test evaluation pairs.} We show the performance of 4 popular state-of-the-art speaker recognition models. $\dagger$ Cleaned versions of these evaluation pairs from the VoxCeleb1 dataset. *VoxSRC2020 validation set\footref{refnote}.}}
\label{table:results}
\vspace{-2mm}
\end{table}
\vspace{-2mm}
\subsection{Baseline models}
We compare four baseline models to investigate the performance on our dataset. All models are trained on the VoxCeleb2 dev set. 

\noindent\textbf{1. I-vector}
\cite{dehak2010front}: Following the Kaldi~\cite{povey2011kaldi} VoxCeleb recipe v1, we extract 400D features, followed by Probabilistic LDA (PLDA) scoring.

\noindent\textbf{2. X-vector}
\cite{snyder2018x}: Following the  Kaldi~\cite{povey2011kaldi} VoxCeleb recipe v2, we train a x-vector model with a PLDA back-end to extract 512D features.

\noindent\textbf{3. Thin ResNet-34}
~\cite{cai2018exploring}: Consists of 34 residual blocks (one-fourth of the channel dimensions of original ResNet-34~\cite{he2016deep}), with self-attentive pooling. ~\cite{chung2020defence} trains this network with an angular variant of the prototypical loss. We use the pre-trained model which is publicly available~\cite{chung2020defence}. \\
\noindent\textbf{4. Thick ResNet-34}
~\cite{heo2020clova}: This has double the number of channels in Thin ResNet-34 and uses attentive statistical pooling to model higher order statistics such as standard deviation. The model is trained with both angular prototypical loss and vanilla softmax loss to improve performance. We use the pre-trained model which is publicly available~\cite{chung2020defence}. This model currently represents the state-of-the-art on the VoxCeleb1 test sets.
512D features are extracted for both thick and thin ResNet-34 architectures. 


\begin{table}[]
\footnotesize
\centering
\begin{tabular}{ccc}
\toprule
\textbf{Evaluation set} & Top1 accuracy & Top5 accuracy \\
\midrule 

VoxCeleb1-test$\dagger$     &    89.47\%   & 97.38\%\\ 
\VoxMovieTest    &  52.23\%     & 73.31\%  \\

\bottomrule
\end{tabular}
\pretabcapt
\caption{\small{Identification results for 485 identities on the \VoxMovieTest\ set (out of domain) and on the VoxCeleb1 test set (same domain). Note how performance drops steeply on the out of domain test set. $\dagger$ VoxCeleb1 test set for identification. }}
\label{identification}
\vspace{-2mm}
\end{table}

\begin{table}[]
\footnotesize
\centering
\begin{tabular}{ccccc}
\toprule
\textbf{Eval set} & Baseline & FT & S-norm & FT + S-norm \\
\midrule
E-1 & 6.09 & 5.76 & 5.89 & \bf{5.66}\\
E-2 & 7.40 & 7.10& 7.18 &\bf{7.03} \\
E-3 & \bf{7.5}0 & 8.38& 8.16 &8.48\\
E-4 & 9.23 & 7.37 & 8.03 &\bf{7.19}\\
E-5 & 10.47& 9.55& 10.15 & \bf{9.35} \\ \midrule
E-3a & \textbf{1.15} & 1.53 & 1.29 & 1.58 \\
E-3b & 0.87 & 0.97 & \textbf{0.68} & 0.98 \\
E-3c & \textbf{7.72} & 9.90 & 9.64 & 10.23 \\
\bottomrule
\end{tabular}
\pretabcapt
\caption{\small{\textbf{Domain transfer results for Thick ResNet-34.} EER(\%) is reported. \textbf{FT:} Fine-tuning on the \VoxMovieTrain\ set. \textbf{S-norm:} Score-normalisation.} }
\vspace{-4mm}
\label{domain transfer}
\end{table}

\vspace{-2mm}
\subsection{Domain Transfer}
In this section, we implement two common domain transfer methods using the \textit{small} amount of data provided in the \VoxMovieTrain\ set. \\
\noindent\textbf{Fine-tuning on a small amount of target domain data.} 
We fine-tune the pretrained Thick-ResNet34 with data from the \VoxMovieTrain\ set and the VoxCeleb2 dev set (overlapping speakers only). 
To decrease the domain gap between the datasets, we always pick one utterance from \datasetname\ and another from VoxCeleb2 to form positive pairs in each mini-batch. This forces the model to decrease the distance between embeddings from the same speaker's utterances, hence reducing the domain gap during training. The network is trained with angular prototypical loss~\cite{chung2020defence}, for 500 epochs using Adam with learning rate of 1e-5. Only the last fully-connected layer is fine-tuned while weights of other layers are fixed.



\noindent\textbf{Score Normalisation.} 
~\cite{matejka2017analysis} introduces various score normalisation techniques for test conditions with diverse domains. A \textit{cohort} is used to estimate the amount of shift and scale for normalisation, allowing robust threshold setting. We experiment with the Z-norm, T-norm and S-norm, and find the best performance to be using the S-norm (Sec.2 in ~\cite{matejka2017analysis}). We use the \VoxMovieTrain\  set as the cohort (speakers in the \VoxMovieTrain\ and (Test) sets are disjoint, which fits the assumption of cohort selection).


\vspace{-1.5mm}
\section{Results}
\vspace{-1mm}
\label{sec:Results}
\noindent\textbf{Verification using Baseline Models.}
The results for the baseline models on the different verification tasks are given in Table~\ref{table:results}. The change of domain in the \datasetname\ evaluation sets offers a significant challenge -- the Thick ResNet-34 which achieves an impressive 1.05 EER on the VoxCeleb1 test set can only achieve 6.09 EER on the least challenging set, E-1. When comparing eval sets that share a positives or negatives source (see Table~\ref{eval_stats}), several conclusions are made: (1) Verifying the same speaker with cross domain utterances (\CrossDomain\ - E-4) is harder than with utterances from the same movie (\VoxMovieDomainsamemovie\ - E-1). Interestingly, this shows that the change in an actor's voice from a calm interview setting in \VoxCelebDomain\ to strong emotions, accents and different background noise in \VoxMovieDomain, is more challenging for speaker verification than the differing emotions in an actor's voice within the same movie at different points in the story-line. Furthermore, the same speaker from different movies (\VoxMovieDomaindifferentmovie\ - E-5) is harder still, showing that an actor's voice changes most between different movies. (2) Negative pair verification is hardest when the negatives are both taken from the unseen domain, \VoxMovieDomain\ (E-4). This is more difficult than when they are taken from \VoxCelebDomain\ (E-3) or \CrossDomain\ (E-2), which are of roughly equal difficulty to the Thick ResNet-34. This is largely because the baseline models were trained on \VoxCelebDomain, so any source of negatives or positives exclusively from that domain will be less challenging. \\
\noindent\textbf{Identification.} 
We use the Thin ResNet-34 model for identification. Table~\ref{identification} shows identification accuracy on both domains. As expected, the identification accuracy drops significantly from VoxCeleb1-test to \VoxMovieTest\ by 37.24\% (top1\% acc.). Anne Hathaway is one of the hardest to identify (top1 acc. 29.17\% - down from 90\% in VoxCeleb1-test), whereas Samuel L. Jackson is easier (acc. 100\% drops to 72.88\% in movies). This may be due to Hathaway's multiple accents in her movies (Fig.~\ref{fig:teaser}). \\ 
\noindent\textbf{Domain Transfer.}
We report the results in Table~\ref{domain transfer}. Fine-tuning with \VoxMovieTrain\ reduces the EER on most of the evaluation sets. The largest improvement is seen in evaluation sets that showed the worst performance in baseline experiments, namely by 2.04\% in E-4 and 1.12\% in E-5. Both E-4 and E-5 source negatives from \VoxMovieDomain\ , showing that fine-tuning and score normalisation work well when transferring existing models from \VoxCelebDomain\ to \VoxMovieDomain. E-3 on the other hand, which sources negatives from \VoxCelebDomain\ shows performance degradation. We conclude that fine-tuning on \VoxMovieDomain\ degrades the performance on negative pairs only from \VoxCelebDomain. To verify this conclusion, we introduce three new evaluation sets, E-3a (positives: \VoxCelebDomain , negatives: \VoxCelebDomain), E-3b (positives: \VoxCelebDomain , negatives: \CrossDomain), E-3c (positives: \VoxMovieDomain , negatives: \VoxCelebDomain) in Table~\ref{domain transfer}.  The fine-tuned model becomes worse at verification in \VoxCelebDomain\ (as shown by the degradation of E-3a). Performance on negatives from \VoxCelebDomain\ contribute most to this (as shown by the degradation of E-3c, relative to E-3b).

\vspace{-1.5mm}

\vspace{-1mm}
\section{Conclusion}
\vspace{-1mm}
\label{sec:conclusions}

In this paper, we provide a novel speaker recognition dataset from movies called \datasetname\ which contains almost 9,000 utterances with diverse emotion, accents and background conditions. We demonstrate that state-of-the-art models trained on interview data from VoxCeleb degrade significantly on cross-domain evaluation sets from \datasetname\, and while simple domain adaptation techniques boost performance, there is still large room for improvement. We hence encourage the research community to develop new methods and systems for this challenging new domain. 

\vspace{-2mm}
\newpara{Acknowledgements.}
We are grateful to Max Bain for his help with the CMD dataset. AB is funded by an EPSRC DTA Studentship, JH by a Global Korea Scholarship, and AN by a Google PhD Fellowship. This work is supported by the EPSRC Programme Grant Seebibyte EP/M013774/1.
\footnotesize
\bibliographystyle{IEEEtranS}
\bibliography{shortstrings,vgg_local,mybib,vgg_other}
\normalsize

\end{document}